# Design and Practice of the Regulation in Speed with Flywheel

[1]Su Wei[a *], [1]Hao Zhuonan[b] and [1]Zhou Huiming[c]
[1] School of Mechanical Engineering, Beijing Institute of Technology, Beijing, China
[a]email, [b]bit1120150799@sina.com, [c]18874207497@163.com

ABSTRACT – This paper aims at the teaching contents of "Fluctuation and Regulation in Speed of Machines" for students, explores the effect of the flywheel on the speed regulation of the mechanism system and its influencing factors, designs an experimental device with applied value, and develops a comparatively complete experimental scheme by contrasting experimental principles. Using this experimental device, we can in-depth study of the effect of the flywheel on the speed of different drive shafts, and deepen students' understanding and mastery of knowledge about regulation in speed.

KEYWORDS - Speed fluctuation regulation. Mechanical theory. Flywheel. Experimental design

[1]The periodic speed fluctuation of the mechanism system is ubiquitous in the running process. If a large speed fluctuation occurs in the entire cycle, the production process will be deteriorated to varying degrees, and the machine efficiency and product quality will be reduced. Therefore, effective measures must be taken to control the magnitude of fluctuations in speed and reduce the harm caused by speed fluctuations. At present, the main method for adjusting the periodic fluctuation in speed is to install a flywheel on the rotating components of the mechanical system. The regulation in speed is an important content in the teaching of "Mechanical Theories".

## 1 Experimental Design Goals

[2-4]In the current course teaching, the calculation formulas of the coefficient of fluctuation in speed, flywheel inertia moment and equivalent moment of inertia and the calculation method of the maximum increment and decrement work lack the practical application process, and the students' understanding of this part is not in place. In the teaching process of the "Mechanical Theories" course, the unity of conclusions and processes should be emphasized, and the cognitive process should be implemented with corresponding demonstration and verification experiments. However, at present, the recognition of regulation in speed only stays in the content of textbooks. There is no perfect experimental device to enable students to apply theory to practice. This paper designs a complete set of experimental equipment, measuring the parameters of the rotating parts during system operation, analysing the effect of regulation in speed, and formulates feasible experimental schemes to complete the verification experiments. [5]Combining the theory and practice, it can not only enhance students' problem analysis ability, but also improve the effect of practical teaching process.

## 2 Theory of Experiment

[6]Periodic speed fluctuation can be adjusted by installing a flywheel with a large moment of inertia. As the speed increases, the moment of inertia of the flywheel prevents its speed from increasing. The flywheel can store energy and limit the increase in maximum angular speed. When the speed reduced, the moment of inertia of the flywheel prevents its speed from decreasing. The flywheel can release energy and limit the decrease of the minimum angular speed. Thereby achieve the purpose of adjusting the speed fluctuation.

Assuming that the number of moving components in the mechanical system is n, the mass of each component is $m_i$, the relative moment of inertia of each mass center is $J_{Si}$, the velocity of each mass center is $v_{Si}$, and the angular velocity of each component is $\omega_i$, then it can be obtained from the principle of equivalence. When the member rotating at the angular velocity $\omega$ around the axis is selected as the equivalent member, the general expression of the equivalent moment of inertia $J_e$ can be obtained as:

$$J_e = \sum_{i=1}^{n} m_i \left(\frac{v_{S_i}}{\omega}\right)^2 + \sum_{i=1}^{n} J_{S_i} \left(\frac{\omega_i}{\omega}\right)^2$$

The average angular speed commonly used in the project indicates the angular speed of the mechanical operation. The difference of the angular speed can reflect the absolute amount of speed fluctuation during the operation cycle of the machine. The ratio of the maximum and minimum angular speed to the average angular speed represents the degree of the coefficient of fluctuation in speed of the mechanical operation. It is expressed by δ. Obviously, in one cycle, when the mechanical speed rises from maximum angular to minimum angular, the external power does the greatest amount of surplus work (or deficit work) to the system, which is called maximum difference between increment and decrement work.

## 3 Experimental Design and Implementation

### *3.1 Experimental Platform Overall Design*

The entire system consists of four parts: power unit, transmission unit, stamping unit and control unit, as shown in Figure 1.

The main body of the power unit is an electric motor. The power of the motor is connected to the stamping unit via a transmission device. The transmission unit includes a coupling that connects the motor power, a V-belt pulley, a gear set, etc. The power transmission terminal is stamping unit. There are three transmission shafts with significant speed difference, whose end can be equipped with different flywheels and the measuring devices respectively. The real-time speed measurement and display on each axis is achieved through the control unit. Optical encoders are installed on each axis, and the speed signal is transmitted to the MCU through the input capture function. The MCU processes the signals to obtain the real-time rotational speed of the measured axis and displays it on the screen. Reveal the speed fluctuation of the system.

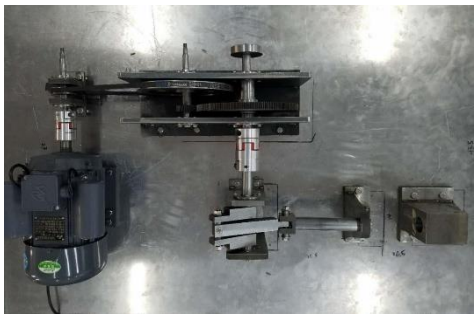
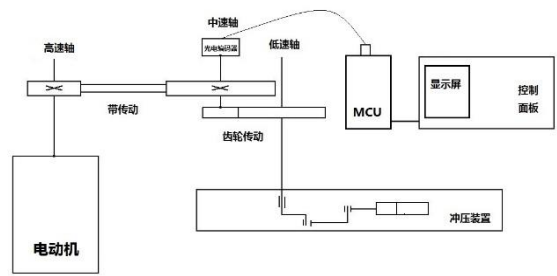

**Fig1** Experiment bench　　　　　　　　　**Fig2** Structure Composition

In order to make the experimental system drive simple and clear and can produce the required experimental results, taking into account the safety and economy, using the structure shown in Figure 2, the power device selection of low-power one-way motor, transmission gear selection - V belt composite gearbox, stamping device selection bias crank slider mechanism. In order to produce more obvious effect of , the end of the stamping device is connected with the spring, so that the crank-slider mechanism presses the spring and converts part of the kinetic energy of the slider into the potential energy of the spring, thereby causing the speed of the mechanical system to change. According to the purpose of the experiment, the three shafts of the transmission device are connected with the flywheel and the speed measuring device is connected with the thimble shaft coupling, so that the rotation speed is synchronized, which is convenient for disassembling and displacement. Given the actual operating conditions of the system, there is no need for large actual impact pressure. The structural parameters of the stamping mechanism (as Figure 3) are set to: $l_1$=70mm，$l_2$=160mm，e=20mm.

### *3.2 Experimental Design Ideas*

In the process of using the flywheel for speed control, the moment of inertia and the installation position of the flywheel will affect the effect of regulation in speed. We know from the existing conclusions:

(1) The effect will be better when the moment of inertia of the flywheel is greater.
(2) The flywheel should be installed on the high-speed shaft.

In order to verify the above conclusions, coefficient of fluctuation in speed was an evaluation index to carry out comparative experiments, including three links below.

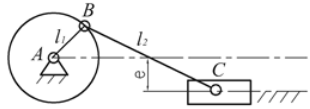
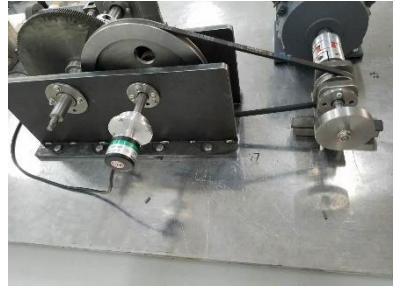

**Fig3** Stamping Mechanism Diagram     **Fig4** Installation Diagram

Experiment 1: Calculation of system equivalent moment of inertia
The entire mechanical system can be regarded as a fixed axis rotation except the slider. Therefore, the rotation speed of the motor is known, and the rotation speed of each axis can be determined by the transmission ratio of each stage. The moment of inertia can be determined by the size parameters of each rotating element. The average speed of the slider can be calculated by the movement characteristics of the crank-slider mechanism. Then we can calculate the equivalent moment of inertia on the medium-speed shaft.

Experiment 2: Effect of the Moment of Inertia of Flywheel on System Speed Variation
The variable is the moment of inertia of the flywheel and three sets of experiments are performed, which are:
a) Without flywheel, the measuring device is fixed on the medium-speed shaft. Then start the equipment. After the device runs stably, measure and record the maximum, minimum and average angular speeds of the medium-speed shaft. Then calculate the coefficient of fluctuation in speed and maximum difference between increment and decrement work of the medium-speed shaft.
b) Install the flywheel A on the high-speed shaft of the transmission. In the same way, calculate the coefficient of fluctuation in speed and maximum difference between increment and decrement work.
c) Install the flywheel B on the high-speed shaft of the transmission. In the same way, calculate the coefficient of fluctuation in speed and maximum difference between increment and decrement work.

Experiment 3: Effect of Flywheel Mounting Position on System Speed Variation
The variable is the flywheel installation location and two sets of experiments are performed, which are:
a) Install the flywheel on the high-speed shaft of the transmission, and install the measuring device on the medium-speed shaft. Measure and record the maximum, minimum and average angular speeds of the medium-speed shaft. Calculate the coefficient of fluctuation in speed and maximum difference between increment and decrement work.
b) Install the flywheel on the low-speed shaft of the transmission. In the same way, calculate the coefficient of fluctuation in speed and maximum difference between increment and decrement work.

### *3.3 Experiment Analysis*

In order to illustrate the design process and test method of this experiment clearly, a example is given below.

In this demonstration experiment, we investigated the influence of the installation of the flywheel on the effect of rotational speed fluctuation of the mechanical system.

The flywheel is mounted on a high-speed shaft and the optical encoder is mounted on a medium speed shaft as shown in Figure 4. Start the motor, wait until the speed is stable, measure and record the speed of the medium-speed shaft, export the data, and use MATLAB to plot the speed curve, as shown in Figure 5. From the figure, the maximum speed is 377.4r/min, the minimum speed is 374.4r/min, and the coefficient of fluctuation in speed is 0.079. Then the flywheel was mounted on the medium-speed shaft as a control group. In the same way, the speed curve of the medium speed shaft was plotted as shown in Figure 6. The maximum speed is 376.2r/min, the minimum speed is 372.6r/min, and the coefficient of fluctuation in speed is 0.024.

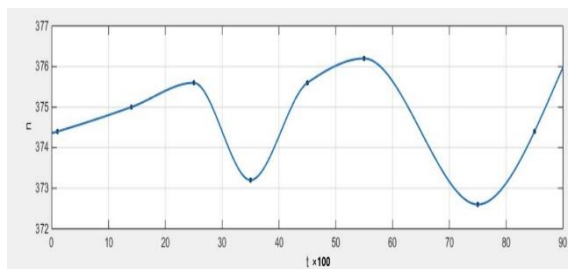 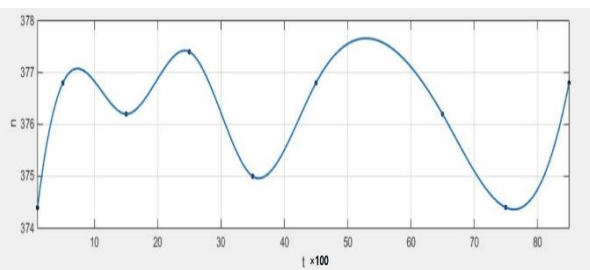

**Fig5** Mid-speed Shaft Speed Curve With Flywheel    **Fig6** Mid-speed Shaft Speed Curve Without Flywheel

Through experiments, it can be seen that when the flywheel is not installed, the speed fluctuation of the mechanical system is obvious, and the coefficient of fluctuation in speed is large. After the flywheel is installed on the high-speed shaft, the coefficient of fluctuation in speed is significantly reduced. Therefore, The flywheel can significantly adjust the speed fluctuation of the mechanical system.

## 4 Summary

This paper quantifies the effect of regulation in speed with the flywheel by a verification experiment, and the measurement result is reliable. Through demonstration experiments, it can be seen that after the installation of the flywheel, the coefficient of fluctuation in speed of the mechanical system drops to 1/3 of the previous one. The results are significant. The experiment is a good reflection of the speed fluctuation of the mechanical system in one cycle, allowing students to grasp the influence factors of the flywheel on the speed control effect. Through practical operation, it deepens the understanding of theoretical knowledge.